\documentclass[preprints,article,accept,moreauthors,pdflatex]{mdpi} 

\firstpage{1} 
\pubvolume{xx}
\issuenum{1}
\articlenumber{5}
\pubyear{2020}
\copyrightyear{2020}
\history{Received: date; Accepted: date; Published: date}

\def\bea{\begin{eqnarray}} \def\eea{\end{eqnarray}}
\def\beq{\begin{equation}} \def\eeq{\end{equation}}
\def\bal#1\eal{\begin{align}#1\end{align}}
\def\bse#1\ese{\begin{subequations}#1\end{subequations}}
\def\non{\nonumber}

\def\text{\mathrm}
\def\be{\beta}
\def\eps{\varepsilon}
\def\la{\Lambda}
\def\tl{\tilde\Lambda}
\def\ms{M_\odot}
\def\mmax{M_\text{max}}
\def\esnm{E_\text{SNM}}

\def\esym{E_\text{sym}}

\def\fm3{\;\text{fm}^{-3}}

\def\rmcore{\epsilon} 

\usepackage{graphicx}
\usepackage{bm}                      
\usepackage{xcolor}                  
\usepackage{amssymb,amsmath}
\usepackage{multirow}

\Title{The Equation of State of Nuclear Matter : from Finite Nuclei to Neutron Stars}

\Author{G. F. Burgio $^{1,\dagger,\ddagger}$\orcidA{}, and I. Vida\~na $^{2,\dagger,\ddagger}$\orcidB{}{}}

\AuthorNames{Fiorella Burgio, and Isaac Vida\~na}

\address{%
$^{1}$ \quad INFN Sezione di Catania, Via S. Sofia 64, I-95123 Catania, Italy; fiorella.burgio@ct.infn.it\\
$^{2}$ \quad INFN Sezione di Catania, Via S. Sofia 64, I-95123 Catania, Italy; isaac.vidana@ct.infn.it}

\corres{Correspondence: fiorella.burgio@ct.infn.it}

\firstnote{These authors contributed equally to this work.}

\abstract{{\it Background.} 
We investigate possible correlations between neutron star observables and properties of atomic nuclei.  Particularly, we explore how the tidal deformability of a 1.4 solar mass neutron star, $M_{1.4}$, and the neutron skin thickness of ${^{48}}$Ca and ${^{208}}$Pb are related to the stellar radius and the stiffness of the symmetry energy.
{\it Methods.} We examine a large set of nuclear equations of state based on phenomenological models (Skyrme, NLWM, DDM) and {\it ab-initio} theoretical methods (BBG, Dirac-Brueckner, Variational, Quantum Monte Carlo). 
{\it Results:} We find strong correlations between tidal deformability and NS radius, whereas a weaker correlation does exist with the stiffness of the symmetry energy. Regarding the neutron skin thickness, weak correlations appear both with the stiffness of the symmetry energy, and the radius of a $M_{1.4}$. {\it Conclusion.} The tidal deformability of a $M_{1.4}$ and the neutron-skin thickness of atomic nuclei show some degree of correlation with nuclear and astrophysical observables, which however depends on the ensemble of adopted EoS.}

\keyword{neutron star; equation of state; many-body methods of nuclear matter; neutron skin thickness; GW170817.}

\begin{document}
{\section{Introduction}
\label{sec:intro}}

The equation of state (EoS) of isospin asymmetric nuclear matter plays a major role in many different realms of modern physics, being the fundamental ingredient for the description of heavy-ion collision dynamics, nuclear structure, static and dynamical properties of neutron stars (NS), core-collapse supernova and binary compact-star mergers \cite{oertel,2018Chap6}. In principle, it can be expected that in heavy ion collisions at large enough energy nuclear matter is compressed at density a few times larger than the nuclear saturation density, and that, at the same time, the two collision partners produce flows of matter, which should be connected with the nuclear EoS. In the physics of compact objects, the central density likely reached in the inner core of a NS may reach values up to one order of magnitude larger than the saturation density, and this poses several theoretical problems because a complete theory of nuclear interactions at arbitrarily large values of density, temperature and asymmetry, should in principle be derived from the quantum chromodynamics (QCD), and this is a very difficult task which presently cannot be realised.
Therefore, theoretical models and methods of the nuclear many-body theory are required to build the EoS, which has to be applied and tested in terrestrial laboratories for the description of ordinary nuclear structure, and in astrophysical observations for the study of compact objects. 

Among possible observables regarding NS, the mass and radius are the most promising since they encode unique information on the EoS at supranuclear densities. Currently the masses of several NSs are known with good precision \cite{mass,demorest2010,fonseca2016,heavy2,cromartie}, but the information on their radii is less accurate \cite{ozel16,gui2013}. The recent observations of NICER \cite{nicer3} have reached a larger accuracy for the radius, but future planned missions like eXTP \cite{2019Watts} will allow us to statistically infer NS mass and radius to within a few percent. 

A big step forward is represented by the recent detection by the Advanced LIGO and Virgo collaborations of gravitational waves emitted during the GW170817 NS merger event \cite{merger,mergerl,mergerx}. This has provided important new insights on the mass and radii of these objects by means of the measurement of the tidal deformability \cite{hartle,flan}, and allowed to deduce upper and lower limits on it \cite{mergerl,radice}.

In this paper we analyze the constraints on the nuclear EoS obtained from the analysis of the NS merger event GW170817, and try to select the most compatible EoS chosen among those derived from both phenomenological and ab-initio theoretical models. We also examine possible correlations among properties of nuclear matter close to saturation with the observational quantities deduced from GW170817 and nuclear physics experiments. In particular, we concentrate on the tidal deformability of NS, and the neutron skin thickness in finite nuclei, thus connecting astrophysical observables with laboratory nuclear physics. 

The paper is organized as follows. In Sect.\ \ref{sec:2} we give a schematic overview of NS phenomenology, whereas in Sect.\ \ref{sec:3} we explain the role of the equation of state in determining the main properties of NS, and illustrate the ones we adopt in the present study. The experimental constraints on the nuclear EoS are presented in Sect.\ \ref{sec:3.1} whereas the astrophysical ones are discussed in Sect.\ \ref{sec:3.2}. A brief overview of different EoS of $\beta$-stable matter is given in Sect.\ \ref{sec:4}, along with numerical results. In Sect.\ \ref{sec:5} we briefly discuss the NS tidal deformability, and its connection to the neutron skin thickness in Sect.\ \ref{sec:6}. Conclusions are drawn in Sect.\ \ref{sec:7}.

\section{Neutron stars in a nutshell}
\label{sec:2}

Neutron stars are a type of stellar compact remnant that can result from the gravitational collapse of an ordinary star with a mass in the range $8-25M_\odot$ (with $M_\odot \approx 2\times 10^{33}$g the mass of the Sun) during a Type II, Ib or Ic supernova event. A supernova explosion will occur when the star has exhausted its possibilities for energy production by nuclear fusion. Then, the pressure gradient provided by the radiation is not sufficient to balance the gravitational attraction, becoming the star unstable and, eventually,   collapsing. The inner regions of the star collapse first and the gravitational energy is released and transferred to the outer layers of the star blowing them away. 

\begin{figure}[t]
\hspace{10mm}
\centerline{\includegraphics[scale=0.45]{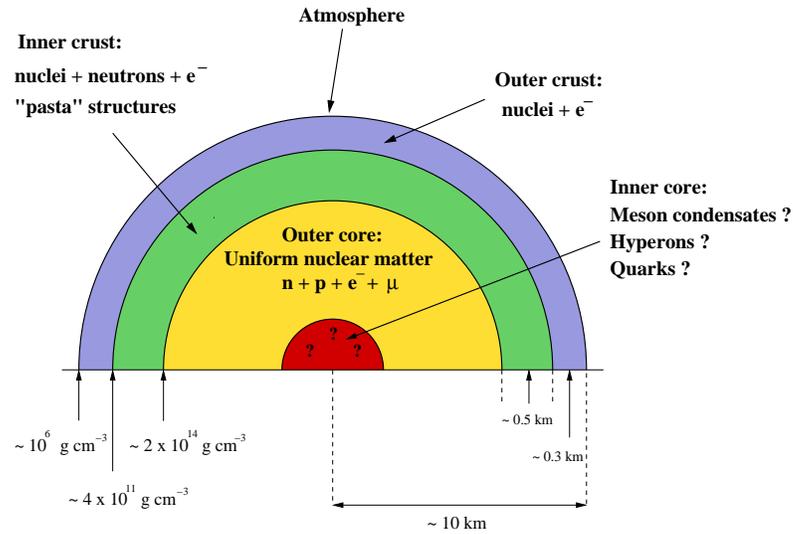}}
\vspace{4mm}
\caption{A schematic cross section of a neutron star illustrating the various regions discussed in the text. The different regions shown are not drawn on scale.
}
\label{f:NSXS}
\end{figure}

NS are supported against gravitational collapse mainly by the neutron degeneracy pressure and may have masses in the range $M\sim 1-2M_\odot$ (with $M_\odot \approx 2\times 10^{33}$g the mass of the Sun) and radii of about $10-12$ km. A schematic cross section the predicted ``onion"-like structure of the NS interior is shown in Fig.\ \ref{f:NSXS}. At the {\it surface}, densities are typically $\rho < 10^{6}$ g/cm$^3$. The {\it outer crust}, with densities ranging from $10^{6}$ g/cm$^3$ to $10^{11}$ g/cm$^3$ is a solid region where heavy nuclei, mainly around the iron mass number, is a Coulomb lattice coexist in $\beta$-equilibrium ({\it i.e.,} in equilibrium with respect to weak interaction processes) with an electron gas. Moving towards the center the density increases 
and the electron chemical potentials increases and the electron capture processes on nuclei
\begin{equation}
e^- + ^AZ \rightarrow ^A(Z-1) + \nu_e \ ,    
\end{equation}
opens and the nuclei become more and more neutron-rich. At densities of $\sim 4.3\times 10^{11}$ g/cm$^3$ the only available levels for the neutrons are in the continuum and they start to ``drip out" of the nuclei. We have then reached the {\it inner crust} region, where matter consist of a Coulomb lattice of very neutron-rich nuclei together with a superfluid neutron gas and an electron gas. In addition, due to the competition between the nuclear and Coulomb forces, nuclei in this region lose their spherical shapes and presenyt more exotic topologies (droplets, rods, cross-rods, salabs, tubves, bubbles) giving rise to what has been called  ``nuclear pasta" phase \cite{pasta}. At densities of $\sim 10^{14}$ g/cm$^3$ nuclei start to dissolve and one enters the {\it outer core}. In this region matter is mainly composed of superfluid neutrons with a smaller concentration of superconducting protons and normal electrons and muons. In the deepest region of the star, the {inner core}, the density can reach values of $\sim 10^{15}$ g/cm$^3$. The composition of this region, however, is not known, and it is still subject of speculation. Suggestions range from a hyperonic matter, meson condensates, or deconfined quark matter.  

The observation of NS requires different types of ground-based and on-board telescopes covering all bands of the electromagnetic spectrum. Radio observations are carried out with ground-based antennas located in different places of the Earth. Three examples of these radio telescopes are the {\it Arecibo radio telescope} in Puerto Rico, the {\it Green Bank Observatory} in West Virginia, and the {\it Nan\c{c}ay decimetric radio telescope} in France. Observations in the near infrared and the optical bands can be performed with the use of
large ground-based telescopes such as the {\it Very Large Telescope} (VLT) in the Atacama desert in Chile. The {\it Hubble-Space Telescope} (HST) can be used used to cover the optical and ultraviolet regions. Observations in the extreme ultraviolet, X-ray and $\gamma$-ray require the use of space observatories such as the {\it Chandra X-ray Observatory} (CXO), the {\it X-ray Multi Mirror} (XMM-Newton) and the {\it Rossi X-ray Timing Explorer} (RXTE) in the case of X-ray observations; and the {\it High Energy Transient Explorer} (HETE-2), the {\it International Gamma-Ray Astrophysics Lboratory} (INTEGRAL) and the {\it Fermi Gamma-ray Space Telescope} (FGST), in the case of $\gamma$-ray ones.

Information on the properties of NS additional to that obtained from the observation of their electromagnetic radiation can be provided from the detection of the neutrinos emitted during the supernova explosion that signals the birth of the star. Examples of neutrino observatories are: the under-ice  {\it IceCube} observatory placed in the South Pole; the under-water projects ANTARES ({\it Astronomy with a Neutrino Telescope and Abyss environmental REsearch}) and the future KM3NET ({\it Cubic Kilometre Neutrino Telescope}) in the Mediterranean sea; and the underground observatories SNO ({\it Sudbury Neutrino Observatory}) located 2100 meters underground in the Vale's Creighton Mine in Canada, and the {\it Kamioka} observatory placed at the Mozumi Mine near the city of Hida in Japan. 

The detection of gravitational waves, originated during the coalescence of two NS as in the GW170817 event recently detected by the Advanced LIGO and Advanced VIRGO collaborations \cite{merger,mergerl,mergerx} or from the oscillation modes of NS, represents nowadays a new way of observing these objects and constitutes a very valuable new source of information. In particular, observations of NS mergers can potentially provide stringent constraints on the nuclear EoS by comparing model predictions with the precise shape of the detected gravitational wave signal. The interested reader is referred to Ref. \ \cite{blaschke19} for a recent review on this hot and exciting topic.

\section{The nuclear equation of state}
\label{sec:3}

The theoretical description of nuclear matter under extreme density conditions
is a very challenging task. Theoretical predictions in this regime are diverse, ranging from purely nucleonic matter with high neutron-proton asymmetry, to baryonic strange matter or a quark deconfined phase of matter. In this work we adopt a conventional description by assuming that the most relevant degrees of freedom are nucleons.  Theoretical approaches to determine the nuclear EoS can be classified in two categories: phenomenological and microscopic ({\it ab-initio}).

 Phenomenological approaches, either non-relativistic or relativistic, are based on effective interactions that are frequently built to reproduce the properties of nuclei \cite{stone}. Skyrme interactions \cite{skyrmea,skyrmeb} and relativistic mean-field (RMF) models \cite{rmfbb,rmfa} are among the most used ones. Many of such interactions are built to describe finite nuclei in their ground state, {\it i.e.} close to the isospin symmetric case and, therefore, predictions at high isospin asymmetries should be taken with care. For instance, most Skyrme forces are, by construction, well behaved close to nuclear saturation density $\rho_0 \approx 0.15-0.16$ fm$^{-3}$ and moderate values of the isospin asymmetry, but predict very different EoS for pure neutron matter, and therefore give different predictions for NS observables. In this work we use the 27 Skyrme forces that passed the restrictive tests imposed by Stone {\it et al.} in Ref.\ \cite{stone} over almost 90 existing Skyrme parametrizations. These forces are: GS and Rs \cite{gsrs}, SGI \cite{sgi}, SLy0-SLy10 \cite{sly} and SLy230a \cite{sly230a1,sly230a2} of the Lyon group, the old SV \cite{sv}, SkI1-Sk5 \cite{ski} and SkI6 \cite{ski6} of the SkI family, SkMP \cite{skmp}, SkO and SkO' \cite{sko}, and SkT4 and SkT5 \cite{skt}.
 
Similarly, relativistic mean-field models are based on effective Lagrangian densities where the interaction between baryons is described in terms of meson exchanges. The couplings of nucleons with mesons are usually fixed by fitting masses and radii of nuclei and the bulk properties of nuclear matter, whereas those of other baryons, like hyperons, are fixed by symmetry relations and hypernuclear observables. In this work we consider two types of RMF models: models with constant meson-baryon couplings described by the Lagrangian density of the nonlinear Walecka model (NLWM), and models with density-dependent couplings [hereafter referred to as density-dependent models (DDM)]. In particular, within the first type, we consider the models GM1 and GM3 \cite{gm}, TM1 \cite{tm1}, NL3 and NL3-II \cite{nl3}, and NL-SH \cite{nlsh}. For the DDM, we consider the models DDME1 and DDME2 \cite{ddme}, TW99 \cite{tw99}, and the models PK1, PK1R and PKDD of the Pekin group \cite{pk}.

Microscopic approaches, on other hand, are based on realistic two- and three-body forces that describe nucleon scattering data in free space and the properties of the deuteron. These interactions are based on meson-exchange theory \cite{bonn1,m3} or, very recently, on chiral perturbation theory \cite{xft1,xft2,xft3,xft4}. Then one has to solve the complicated many-body problem \cite{baldo1999} in order to obtain the nuclear EoS. The main difficulty is the treatment of the short-range repulsive core of the nucleon-nucleon interaction. Different many-body approaches have been devised for the construction of the nuclear matter EoS, e.g.,  the Brueckner--Hartree--Fock (BHF) \cite{bbg} and the Dirac--Brueckner--Hartree--Fock (DBHF) \cite{dbhf1,dbhf2,dbhf3} theories, the variational method \cite{apr1998}, the correlated basis function formalism \cite{cbf}, the self-consistent Green's function technique \cite{scgf1,scgf2}, the $V_{\mbox{low}\,k}$ approach \cite{vlowk} or Quantum Monte Carlo techniques \cite{qmc1,qmc2}. 

As far as the microscopic approaches are concerned, in this paper we adopt several BHF EoS 
based on different nucleon-nucleon potentials, namely the Bonn B (BOB) \cite{bonn1,bonn2}, the Nijmegen 93 (N93) \cite{nij1,nij2}, and the Argonne $V_{18}$ (V18) \cite{v18}. In all those cases, the two-body forces are supplemented by nucleonic three-body forces (TBF), which are needed
in order to reproduce correctly the saturation properties of nuclear matter.
Currently a complete ab-initio theory of TBF is not available yet,
and therefore we adopt either phenomenological or microscopic models
\cite{glmm,uix3,zuo1,tbfnij}. The microscopic TBF employed in this paper are
described in detail in Refs.~\cite{tbfnij,li08}, whereas a phenomenological approach based on the Urbana model \cite{uix1,uix2,uix3}, is also adopted. In this case the corresponding EoS is labelled UIX in Table~\ref{t:sat}.  
Within the same theoretical framework,
we also studied an EoS based on a potential model which includes
explicitly the quark-gluon degrees of freedom, named FSS2
\cite{2014PhRvL.113x2501B,2015PhRvC..92f5802F}.
This reproduces correctly the saturation point of symmetric matter
and the binding energy of few-nucleon systems
without the need of introducing TBF. In the following we use two different EoS versions labelled respectively as FSS2CC and FSS2GC. Moreover, we
compare these BHF EoSs with the often-used results of the
Dirac-BHF method (DBHF) \cite{dbhf3},
which employs the Bonn~A potential,
the APR EoS \cite{apr1998}
based on the variational method and the $V_{18}$ potential, and a parametrization of a recent Auxiliary Field diffusion Monte Carlo (AFDMC) calculation of Gandolfi {\it et al.} given in Ref.\ \cite{afdmc}. 

The above mentioned methods provide EoSs for homogeneous nuclear matter,
$\rho > \rho_t \approx 0.08\,\text{fm}^{-3}$.
For the low-density inhomogeneous part
we adopt the well-known Negele-Vautherin EoS
\cite{NV} for the inner crust in the medium-density regime
($0.001\,\text{fm}^{-3} < \rho < 0.08\,\text{fm}^{-3}$),
and the ones by Baym-Pethick-Sutherland \cite{bps}
and Feynman-Metropolis-Teller \cite{fey} for the outer crust
($\rho < 0.001\,\text{fm}^{-3}$).

\subsection{Laboratory constraints on the nuclear EoS}
\label{sec:3.1}
 Around saturation density $\rho_0$ and isospin asymmetry $\delta \equiv (N-Z)/(N+Z)=0$
[being $N (Z)$ the number of neutrons (protons)], the nuclear EoS can be characterized by a set of few isoscalar ($E_0, K_0$) and isovector ($S_0, L, K_{sym}$) parameters. These parameters can be constrained by nuclear experiments and are related to the coefficients of a Taylor expansion of the energy per particle of asymmetric nuclear matter as a function of density and isospin asymmetry
\bea
 E(\rho,\delta) &=& \esnm(\rho) + \esym(\rho) \delta^2 \:,
\label{e:ea}
\\
 \esnm(\rho) &=& E_0 + \frac{K_0}{2} x^2 \:,
\\
 \esym(\rho) &=& S_0 + L x + \frac{K_\text{sym}}{2} x^2 \:,
\label{e:ebulk}
\eea
where $x \equiv (\rho-\rho_0)/3\rho_0$, $E_0$ is the energy per particle of symmetric nuclear matter at $\rho_0$, $K_0$ the incompressibility and
$S_0 \equiv E_\text{sym}(\rho_0)$
is the symmetry energy coefficient at saturation.
The parameters $L$ and $K_\text{sym}$
characterize the density dependence of the symmetry energy around saturation.
These parameters are defined as
\bea
 K_0 &\equiv& 9\rho_0^2  \frac{d^2\esnm}{d\rho^2}(\rho_0) \:,
\\
 S_0 &\equiv& \frac{1}{2} \frac{\partial^2E}{\partial\delta^2}(\rho_0,0)
  \:,
\\
 L &\equiv& 3 \rho_0 \frac{d\esym}{d\rho}(\rho_0) \:,
\\
 K_\text{sym} &\equiv& 9 \rho_0^2 \frac{d^2\esym}{d\rho^2}(\rho_0) \:.
\eea
The incompressibility $K_0$ gives the curvature of $E(\rho)$ at $\rho=\rho_0$,
whereas $S_0$ determines the increase of the energy per nucleon
due to a small asymmetry $\delta$.

Properties of the various considered EoSs are listed in Table~\ref{t:sat},
namely, the value of the saturation density $\rho_0$,
the binding energy per particle $E_0$,
the incompressibility $K_0$,
the symmetry energy $S_0$,
and its derivative $L$ at $\rho_0$. 
Measurements of nuclear masses \cite{audi03} and density distributions \cite{vries87} yield $E_0=-16\pm 1$ MeV and $\rho_0=0.14-0.17$ fm$^{-3}$, respectively.  The value of $K_0$ can be extracted from the analysis of isoscalar giant monopole resonances in heavy nuclei, and results of Ref.\ \cite{colo04} suggest  $K_0=240\pm 10$ MeV, whereas in Ref.\ \cite{piekarewicz04} a value of $K=248 \pm 8$ MeV is reported.  Even heavy ion collision experiments point to a  ``soft" EoS, {\it i.e.,} a low value of $K_0$, though the constraints inferred from heavy ion collisions are model dependent because the analysis of the measured data requires the use of transport models \cite{fuchs01}. 
Experimental information on the symmetry energy at saturation $S_0$ and its derivative $L$  can be obtained from several sources such as the analysis of giant \cite{giant} and pygmy \cite{pygmy1,pygmy2} dipole resonances, isospin diffusion measurements \cite{isodif}, isobaric analog states \cite{isob}, measurements of the neutron skin thickness in heavy nuclei \cite{skin1,skin2,skin3,skin4,skin5} and meson production in heavy ion collisions \cite{meson}.  However, whereas $S_0$ is more or less well established ($\approx 30$ MeV), the values of $L$ ($30\,\text{MeV} < L < 87\,\text{MeV}$), and especially those of $K_{sym}$ ($-400\,\text{MeV} < K_\text{sym} < 100\,\text{MeV}$) are still quite uncertain and poorly constrained \cite{tews2017,2017arXiv170402687Z}.
The reason why the isospin dependent part of the nuclear EoS is so uncertain is still an open question, very likely related to our limited knowledge of the nuclear forces and, in particular, to its spin and isospin dependence.

From Table~\ref{t:sat}, we notice that all the adopted EoSs in this work agree fairly well with the empirical values.
Marginal cases are the
slightly too low $E_0$ and $K_0$ for V18,
too large $S_0$ for N93,
and too low $K_0$ for UIX and FSS2GC. We notice that the 
$L$ parameter does not exclude any of the microscopic EoSs, whereas several phenomenological models predict too large $L$ values. 

\begin{table}[!p]
\renewcommand{\arraystretch}{1.15}
\begin{center}
\caption{
Saturation properties predicted by the considered EoSs.
Experimental nuclear parameters are listed for comparison.
See text for details.}
\label{t:sat}
\begin{tabular}{ccccccc}
\hline\hline
 Model class & EoS    & $\rho_0[\fm3]$ & $-E_0$[MeV] & $K_0$[MeV] & $S_0$[MeV] & $L$[MeV] \\
\hline
Skyrme & Gs & 0.158 &  14.68 & 239.34 & 39.55  & 93.55 \\
& Rs & 0.158  & 14.01 &  248.33 &  38.45 &  86.41 \\
& SGI & 0.155 & 14.67 & 265.35 & 34.32 & 63.85 \\
& SLy0 &    0.16   &    15.28   &     226.42  &     34.82    &    45.37 \\
& SLy1 & 0.161    &   15.23    &   233.25  &     36.21   &     48.88 \\
& SLy2 & 0.161    &    15.16     &   234.54  &      36.01  &     48.84 \\
& SLy3 &   0.161    &    15.22   &     232.85  &      35.45     &   45.56 \\
& SLy4 &     0.16   &     15.18   &    232.19  &      35.26   &     45.38 \\
& SLy5 &   0.161  &     15.25  &      232.03    &    36.44  &     50.34 \\
& SLy6 &   0.159  &      15.10  &     230.09 &      34.74 &       45.21 \\
& SLy7 &  0.159     &   15.05   &     233.10  &      36.18   &     48.11 \\
& SLy8 &    0.161 &       15.22  &     233.34   &     34.84    &   45.36 \\
& SLy9 &    0.151    &   14.53   &    228.95 &      37.72 &      55.37 \\
& SLy10 &    0.156    &    14.92   &     231.75  &      35.32   &     39.24 \\
& SLy230a &    0.16 &      15.22   &     229.98 &      35.26  &      43.99 \\
& SV &    0.155   &     14.65 &     304.99  &      42.42  &    96.51 \\
& SkI1 &   0.161  &      15.59 &       233.87   &     51.24  &      160.46 \\
& SkI2 &    0.158   &     14.78   &    245.14  &      43.38  &     105.72 \\
& SkI3 &    0.158   &    14.99   &    259.44 &        44.32 &      101.16 \\
& SkI4 &    0.16 &       15.42   &    238.92  &      34.21  &    59.34 \\
& SkI5 & 0.156     &   14.73       & 257.41    &    49.44  &      129.29 \\
& SkI6 & 0.158   &    14.98    &    243.93   &    41.62  &   81.76 \\
& SkMP &   0.157  &     14.66   &     230.16  &     35.88 &      69.7 \\
& SkO & 0.161    &    15.04  &     228.10 &     38.52  &      79.92 \\
& SkO' & 0.16  &     14.99   &    222.28    &   37.66 &      69.68 \\
& SkT4 & 0.159  &      15.12  &     235.48 &       43.19   &     93.48 \\
& SkT5 & 0.164  &      15.48  &     201.66 &       44.88   &     100.32 \\
\hline 
NLWM & GM1 & 0.153 & 16.34 & 300.28 & 32.49 & 93.92 \\
& GM3 &  0.153 &  16.36 &   240.53 &    32.54 & 89.83 \\
& TM1 &  0.145 & 16.26 &    281.16 &    36.89 & 110.79 \\
& NL3 & 0.148 &  16.24 & 271.54 &   37.4 &  118.53  \\
& NL3-II &  0.149 &  16.26 & 271.74 &  37.70 &  119.71 \\
& NL-Sh &  0.146 &  16.36 &  355.65 &   36.13 & 113.68 \\
\hline
DDM & DDME1 & 0.152 &  16.2 &  244.72 & 33.067 &   55.46 \\
& DDME2 & 0.152 &   16.14 &  250.9 & 32.3 & 51.26 \\
& TW99 & 0.153 & 16.25 & 240.26 &   32.766 & 55.31 \\
& PK1 & 0.148 &  16.27 & 282.7 & 37.64 & 115.88 \\ 
& PK1R & 0.148 &    16.27 &  283.68 &  37.83 &  116.5 \\
& PKDD & 0.149 & 16.27 &    262.19 & 36.79 & 90.21 \\
\hline
 Microscopic & BOB     & 0.170  & 15.4 & 238 & 33.7 & 70  \\
 & V18     & 0.178  & 13.9 & 207 & 32.3 & 67  \\
 & N93     & 0.185  & 16.1 & 229 & 36.5 & 77  \\
 & UIX     & 0.171  & 14.9 & 171 & 33.5 & 61  \\
 & FSS2CC  & 0.157  & 16.3 & 219 & 31.8 & 52  \\
 & FSS2GC  & 0.170  & 15.6 & 185 & 31.0 & 51  \\
 & DBHF    & 0.181  & 16.2 & 218 & 34.4 & 69  \\
 & APR     & 0.159  & 15.9 & 233 & 33.4 & 51  \\
 & AFDMC   & 0.160  & 16.0 & 239 & 31.3 &  60 \\
\hline
& Exp.   & $\sim$ 0.14--0.17 & $\sim$ 15--17 & 220--260 & 28.5--34.9 & 30--87 \\
& Ref.   & \cite{margue2018a} 
        & \cite{margue2018a} 
        & \cite{shlomo,pieka}
        & \cite{lihan,oertel} & \cite{lihan,oertel} \\
\hline\hline
\end{tabular}
\end{center}
\end{table}


\subsection{Astrophysical constraints on the nuclear EoS}
\label{sec:3.2}

The main astrophysical constraints on the nuclear EoS are those arising from the observation of NS. An enormous amount of data on different NS observables have been collected after fifty years of NS observations. These observables include: masses, radii, rotational periods, surface temperatures, gravitational redshifts, quasi-periodic oscillations, magnetic fields, glitches, timing noise and, very recently, gravitational waves. In the next lines we briefly review how masses and radii are measured. Observational constraints derived from the recent observation of the gravitational wave signal from the merger of two NS detected by the Advanced LIGO and Advanced VIRGO collaborations \cite{merger,mergerl,mergerx} will be discussed in detail in Sect.~\ref{sec:5}.

NS masses can be directly measured from observations of binary systems. There are five orbital parameters, also known as Keplerian parameters, which can be precisely measured. They are the projection of the pulsar's semi-major axis ($a_1$) on the line of sight ($x\equiv a_1 \mbox{sin}\, i/c$, where $i$ is inclination of the orbit), the eccentricity of the orbit ($e$), the orbital period ($P_b$), and the time ($T_0$) and longitude ($\omega_0$) of the periastron. With the use of Kepler's Third Law, these parameters can be related to the masses of the NS ($M_p$) and its companion ($M_c$) though the so-called mass function
\begin{equation}
f(M_p,M_c,i)=\frac{(M_c\,\mbox{sin}\,i)^3}{(M_p+M_c)^2}=\frac{P_bv_1^3}{2\pi G}
\label{eq:massfun}
\end{equation}
where $v_1=2\pi a_1\mbox{sin}\,i/P_b$ is the projection of the orbital velocity of the NS along the line of sight. The individual masses of the two components of the system cannot be obtained if only the mass function is determined. Additional information is required. Fortunately, deviations from the Keplerian orbit due to general relativity effects can be detected. The relativistic corrections to the orbit are parametrized in terms of one or more parameters called post-Keplerian. The most significant ones are: the combined effect of variations in the transverse Doppler shift and gravitational redshift around an elliptical orbit ($\gamma$), the range (r) and shape (s) parameters that characterize the Shapiro time delay of the pulsar signal as it propagates through the gravitational field of its companion, the advance of the periastron of the orbit ($\dot \omega$) and the orbital decay due to the emission of quadrupole gravitational radiation ($\dot P_b$). These post-Keplerian parameters can be written in terms of measured quantities and the masses of the star and its companion (see {\it e.g.,} Ref.\ \cite{taylor92} for specific expressions). The measurement of any two of these post-Keplerian parameters together with the mass function $f$ is sufficient to determine uniquely the masses of the two components of the system.

As the reader can imagine NS radii are very difficult to measure, the reason being that NS are very small objects and are very far away from us ({\it e.g.,} the closest NS is the object RX J1856.5-3754 in the constellation Corona Australis which is about 400 light-years from the Earth). That is why there not exist direct measurements of NS radii yet. Nevertheless, it is possible to determine them by using the thermal emission of low-mass X-ray binaries (systems where one of the components is a NS and the companion a less massive object ($M_c < M_\odot $) which can be a main sequence star, a red giant or a white dwarf). The observed X-ray flux ($F$) and estimated surface temperature ($T$) together with a determination of the distance ($D$) of the star, can be used to obtain the radius of the NS through the implicit relation
\begin{equation}
R=\sqrt{\frac{FD^2}{\sigma T^4}\left(1-\frac{2GM}{c^2R}\right)} \ .
\label{ec:radns}
\end{equation}
Here $\sigma$ is the Stefan--Boltzmann constant and $M$ the mass of the NS. The major uncertainties in the measurement of the radius through Eq.\ (\ref{ec:radns}) come from the determination of the temperature, which requires the assumption of an atmospheric model, and the estimation of the distance of the star. However, the analysis of present observations from quiescent low-mass X-ray binaries is still controversial (see {\it e.g.,} Refs. \cite{lat2014,2014Gui}).

We notice that the simultaneous measurement of both mass and radius of the same NS would provide the most definite observational constraint on the nuclear EoS. Very recently the NICER (Neutron Star Interior Composition Explorer) mission has reported a Bayesian parameter estimation of the mass and equatorial radius of the millisecond pulsar PSR J0030+0451 \cite{nicer3}. The values inferred from the analysis of the collected data are $1.34^{+0.15}_{-0.16}M_\odot$ and $12.71^{+1.14}_{-1.19}$ km, respectively. 


\section{EoS for $\beta$-stable matter}
\label{sec:4}

In order to study the structure of the NS core, we have to calculate the 
composition and the EoS of cold, neutrino-free, catalyzed matter. As stated before,
we consider a NS with a core of nucleonic matter
without hyperons or other exotic particles. We require that it contains charge
neutral matter consisting of neutrons, protons, and leptons ($e^-$, $\mu^-$) in
$\beta$-equilibrium, and compute the EoS for charge neutral and $\beta$-stable matter
in the following standard way \cite{shapiro}. The output of the many-body calculation is the energy density of lepton/baryon matter as a function of the different densities $\rho_i$ of the species $i=n,p,e,\mu$ ,
\begin{equation}
 \rmcore(\rho_n,\rho_p,\rho_e,\rho_\mu) 
 = (\rho_n m_n +\rho_p m_p) 
 + (\rho_n+\rho_p) E(\rho_n,\rho_p)
 + \rmcore(\rho_e) + \rmcore(\rho_\mu) \:,
\label{e:epsnn}
\end{equation}
where $m_i$ are the corresponding masses, and $E(\rho_n,\rho_p)$ is the enegy per particle of asymmetric nuclear matter. 
We have used ultrarelativistic and relativistic expressions
for the energy densities of electrons $\rmcore(\rho_e)$ and muons $\rmcore(\rho_\mu)$, respectively \cite{shapiro}.
Since microscopic calculations are very time consuming in the case of these models we have used the parabolic approximation \cite{hypns1,hypns2,bom1,bom2,bom3} of the energy per particle of asymmetric nuclear matter given in Eq.\ (\ref{e:ea}) with the symmetry energy calculated simply as the difference between the energy per particle of pure neutron matter $E(\rho_n=\rho,\rho_p=0)$ and symmetric nuclear matter $E(\rho_n=\frac{\rho}{2},\rho_p=\frac{\rho}{2})$
\begin{equation}
  E_{\rm sym}(\rho)  \approx E(\rho_n=\rho,\rho_p=0) -E(\rho_n=\frac{\rho}{2},\rho_p=\frac{\rho}{2}) \:.
\label{e:sym}
\end{equation}

Once the energy density (Eq.~(\ref{e:epsnn})) is known 
the various chemical potentials can be computed straightforwardly,
\begin{equation}
 \mu_i = {\partial \rmcore \over \partial \rho_i} \:,
\end{equation}
and solving the equations for $\beta$-equilibrium,
\begin{equation}
\mu_i = b_i \mu_n - q_i \mu_e \:,
\end{equation}
($b_i$ and $q_i$ denoting baryon number and charge of species $i$)
along with the charge neutrality,
\begin{equation} 
 \sum_i \rho_i q_i = 0 \:,
\end{equation}
allows one to find the equilibrium composition $\rho_{i}$
at fixed baryon density $\rho$, and finally the EoS,
\begin{equation}
 P(\rmcore) = \rho^2 {d\over d\rho} 
 {\rmcore(\rho_i(\rho))\over d\rho}
 = \rho {d\rmcore \over d\rho} - \rmcore 
 = \rho \mu_n - \rmcore \:.
\end{equation}

Once the EoS of $\beta$-stable matter is known, one can determine the hydrostatical equilibrium configurations just solving the 
Tolman-Oppenheimer-Volkoff (TOV) \cite{shapiro} equations which describe the structure of a non-rotating spherically symmetric star in general relativity:
\begin{eqnarray}
{d P\over d r}&=&- G\, {\rmcore m \over r^2} \left(1 + {P \over \rmcore} \right) \left(1 + {4\pi P r^3\over m } \right) \left(1 - {2 G m \over r }\right)^{-1} \nonumber \\
{d m\over d r}&=&4\pi r^2 \rmcore \,,
\label{eq:OV} 
\end{eqnarray}
\noindent where $G$ is the gravitational constant, $P$ the pressure, $\rmcore$ the energy density,  
$m$ the mass enclosed within a sphere of radius $r$. The TOV equations have an easy interpretation. Consider a spherical shell of matter of radius $r$ and thickness $dr$. The second equation gives the mass in this shell whereas the left hand side of the first one is the net force acting on the surface of the shell by the pressure difference $dP(r)$. The first factor of the right hand side of this equation is the attractive Newtonian force of gravity acting on the shell by the mass interior to it. The remaining three factors result from the correction of general relativity. So the TOV equations express the balance at each $r$ between the internal pressure as it supports the overlying material against the gravitational attraction of the mass interior to $r$. The integration of the TOV equations gives the mass and radius of the star for a given central density. It turns out that the mass of the NS has a maximum value as a function of radius (or central density), above which the star
is unstable against collapse to a black hole. The value of the maximum mass
depends on the nuclear EoS, so that the observation of a mass higher than the
maximum mass allowed by a given EoS simply rules out that EoS.

\begin{figure*}[!htbp]
\centerline{\includegraphics[scale=0.65]{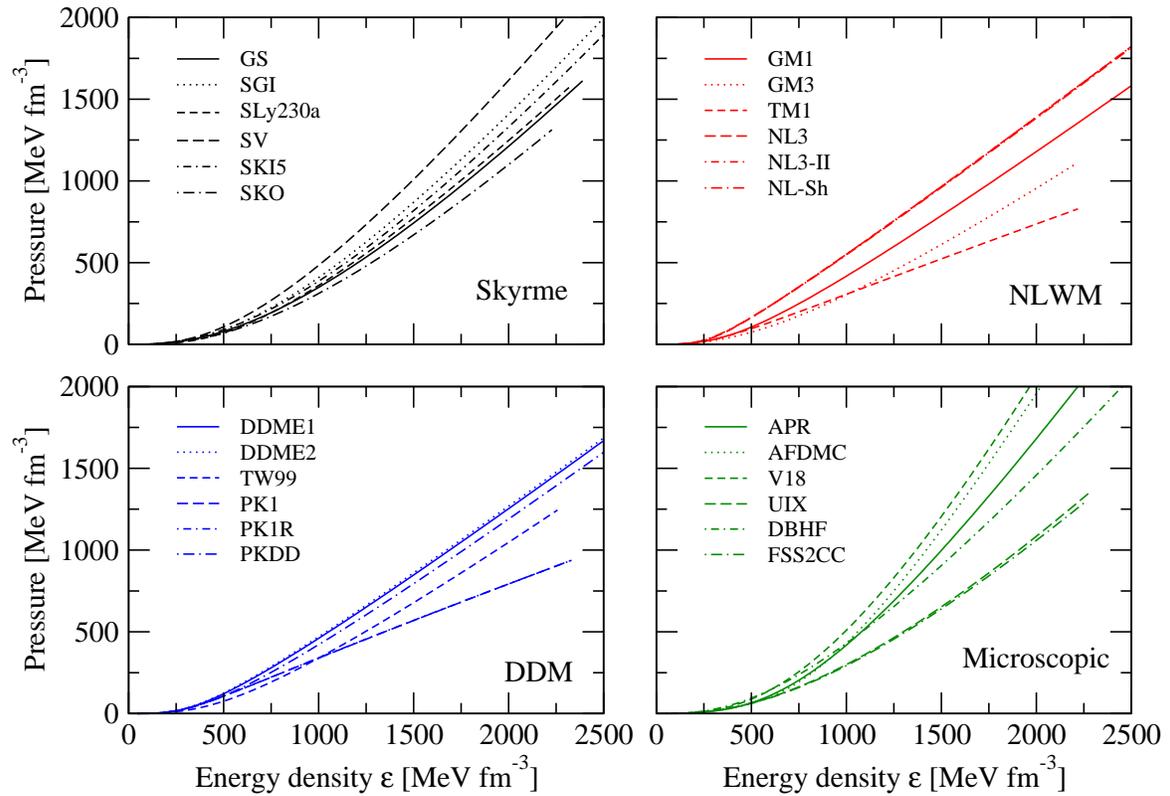}}
\vspace{4mm}
\caption{Equation of state of $\beta$-stable matter for the four model classes reported in Table\ref{t:sat}. 
}
\label{f:fig1}
\end{figure*}

We now turn to the discussion of some results. We display in Fig.\ref{f:fig1} the $\beta$-stable matter EoS obtained for some of the models illustrated in Table \ref{t:sat}, a limited sample of each class being plotted in one single panel. We see that the pressure is a monotonically increasing function of the energy density for all EoS. Each EoS is characterized by a given stiffness, which determines the maximum mass value of a NS: the stiffer the EoS the larger the maximum mass predicted.

\begin{figure*}[!htbp]
\centerline{\includegraphics[scale=0.65]{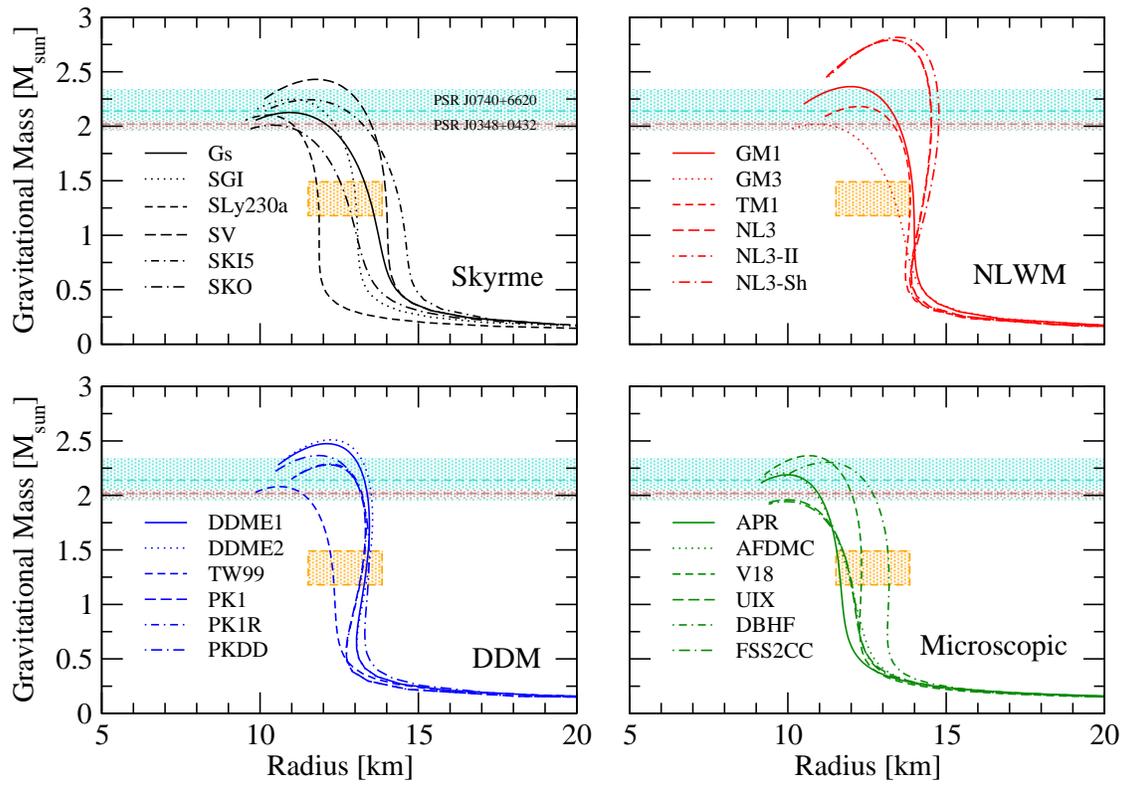}}
\vspace{4mm}
\caption{Mass-radius relation predicted by the different EoS displayed in Fig.\ref{f:fig1}. The observed masses of the millisecond pulsars PSR J0740+6620 \cite{demorest2010} and PSR J0348+0432 \cite{heavy2} are also shown.
The box shows the constraints inferred from the observations reported by the NICER mission \cite{nicer3}. See text for details.
}
\label{f:fig2}
\end{figure*}

The corresponding mass-radius relation is displayed in Fig.\ref{f:fig2}.
The observed trend is consistent with the EoS displayed in Fig.\ref{f:fig1}. 
As expected, when the EoS stiffness increases the NS maximum mass increases as well. 
The considered EoS are compatible with the largest masses observed up to now,
$\mmax>2.14^{+0.10}_{-0.09}$ \cite{cromartie} for the object PSR J0740+6620 (cyan hatched area), and
PSR J0348+0432 \cite{heavy2},
$M_G=2.01 \pm 0.04 \,M_{\odot}$ (red hatched area). We notice that recent analysis of the GW170817 event
indicate also an upper limit of the maximum mass of about
2.2--2.3$\,\ms$
\cite{shiba17,marga17,rezz18,shiba19},
with which most of the models shown in the figure are compatible. The box shows the estimation of the mass ($1.34^{+0.15}_{-0.16}M_\odot$) and equatorial radius ($12.71^{+1.14}_{-1.19}$ km) of the millisecond pulsar PSR J0030+0451 inferred from the Bayesian analysis of the data collected by the NICER mission \cite{nicer3}. Note that this constraint excludes most of the NLWM EoS considered in this work.

\vfill
\section{The neutron star tidal deformability}
\label{sec:5}

Recently the tidal deformability $\lambda$,
or equivalently the tidal Love number $k_2$ of a NS
\cite{hinder2008,hinder2009,hinder2010},
has been recognized to provide valuable information and constraints
on the related EoS, because it strongly depends on the compactness of the object, i.e. $\be\equiv M/R$.
More specifically, the Love number
\bea
 k_2 &=& \frac{3}{2}\frac{\lambda}{R^5} = \frac{3}{2} \be^5 \la
 = \frac{8}{5}\frac{\be^5 z}{F} \:,
\label{e:l}
\\
 && z \equiv  (1-2\be)^2 [2-y_R+2\be(y_R-1)] \:,
\non\\
 && F \equiv  6\be(2-y_R) + 6\be^2(5y_R-8) + 4\be^3(13-11y_R)
\non\\\non
   &&\hskip6mm  +\, 4\be^4(3y_R-2) + 8\be^5(1+y_R) + 3z\ln(1-2\be) \:
\eea
with $\la\equiv\lambda/M^5$,
can be obtained by solving the TOV equations~(\ref{eq:OV}),
along with the following first-order differential equation
\cite{EoS},
\bea
  {dy \over dr} &=& -\frac{y^2}{r} - \frac{y-6}{r-2m} - rQ\:,
\non\\
  && Q \equiv 4\pi \frac{(5-y)\eps+(9+y)P+(\eps+P)/c_s^2}{1-2m/r}
\non\\
  && \phantom{Q \equiv} - \Bigg[ \frac{2(m+4\pi r^3 P)}{r(r-2m)} \Bigg]^2,
\label{e:tov3}
\eea
with the EoS $P(\eps)$ as input,
$c_s^2=d\!P/d\eps$ the speed of sound,
and boundary conditions given by
\beq
 [P,m,y](r=0) = [P_c,0,2] \:,
\eeq
being $y_R\equiv y(R)$,
and the mass-radius relation $M(R)$
provided by the condition $P(R)=0$ for varying central pressure $P_c$.

For an asymmetric binary NS system,
$(M,R)_1+(M,R)_2$, with mass asymmetry $q=M_2/M_1$,
and known chirp mass $M_c$, which characterizes the GW signal waveform,
\beq
 M_c = \frac{(M_1 M_2)^{3/5}}{(M_1+M_2)^{1/5}} \:,
\eeq
the average tidal deformability is defined by
\beq
 \tl = \frac{16}{13}
 \frac{(1+12q)\la_1 + (q+12)\la_2}{(1+q)^5}
\label{e:lq}
\eeq
with
\beq
 \frac{[M_1,M_2]}{M_c} =
 \frac{297}{250} (1+q)^{1/5} [q^{-3/5},q^{2/5}] \:.
\eeq

From the analysis of the GW170817 event \cite{merger,mergerl,mergerx},
a value of $M_c=1.186{+0.001\atop-0.001}\ms$ was obtained,
corresponding to $M_1=M_2=1.36\,\ms$ for a symmetric binary system,
$q=0.73-1$ and $\tl<730$ from the phase-shift analysis of the observed signal.
It turns out that, requiring both NSs to have the same EoS,
leads to constraints $70 < \la_{1.4} < 580$ and
$10.5 < R_{1.4} < 13.3$ km \cite{mergerl} for a 1.4 solar mass NS. 

However the high luminosity of the kilonova AT2017gfo following the NS merger event,
imposes a lower limit on the average tidal deformability, Eq.~(\ref{e:lq}), $\tl>400$, which
was deduced in order to justify the amount of ejected material heavier than $0.05\,\ms$. This constraint could indicate that $R_{1.4}\gtrsim 12\,$km,
which was used in Refs.~\cite{most18,lim18,malik18,drago4}
in order to constrain the EoS.
This lower limit has to be taken with great care and, in fact, it has been recently revised to $\tl \gtrsim 300$ \cite{radice2}, but considered of limited significance
in Ref.~\cite{kiuchi}.

\begin{figure}[!htbp]
\centerline{\includegraphics[scale=0.65]{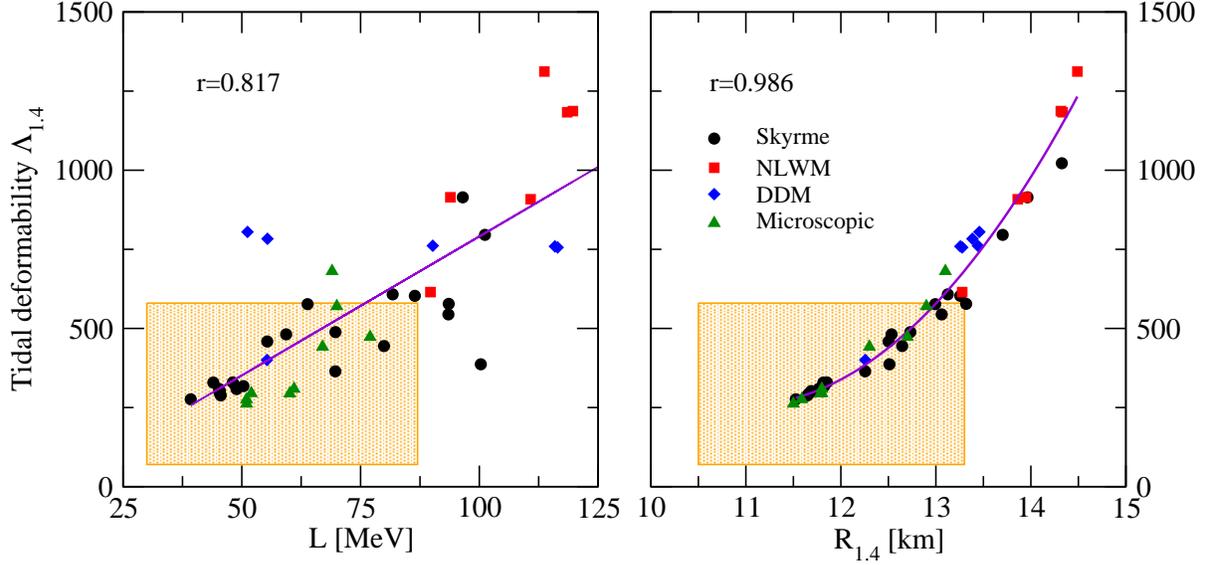}}
\vspace{4mm}
\caption{In the left panel the tidal deformability of a 1.4 solar mass NS $\Lambda_{1.4}$ is plotted vs. the symmetry energy derivative at saturation density $L$, whereas in the right panel it is displayed as a function of the radius of a 1.4 solar mass NS, $R_{1.4}$, for the different EoS shown in Table~\ref{t:sat}. The orange box indicate the experimental and observational constraints on $L$ (see Table \ref{t:sat}), $\la_{1.4}$ and $R_{1.4}$ \cite{mergerl}. The violet line indicates a linear (quadratic) fit of the EoS data. The values of the corresponding correlation factors are also given. See text for details.
}
\label{f:fig3}
\end{figure}

One of the main theoretical issues, following the detection of gravitational waves from NS mergers, regards the possibility of finding correlations between properties of nuclear matter and NS observables \cite{wei_epja}. Along this same line, we 
further explore this issue, and using the set of microscopic EoS and the several Skyrme forces and relativistic models listed in Table~\ref{t:sat}, in the left panel of Fig.~\ref{f:fig3} we show the tidal deformability of a 1.4 solar mass NS as a function the symmetry energy parameter $L$ at saturation density. The orange box shows the constraint on $\la_{1.4}$ inferred from the observational data of the GW170817 event \cite{mergerl} together with the experimental limits of $L$ reported in Table~\ref{t:sat}. We observe some degree of correlation between the tidal deformability and $L$ , for which we can estimate the so-called correlation factor $r$, defined as
\beq
 r(L,\Lambda_{1.4}) = \frac{1}{n-1}
 \frac{\sum_{L} \sum_{\Lambda_{1.4}} (L-\bar{L}) (\Lambda_{1.4}-\bar{\Lambda}_{1.4})}{\sigma_L \sigma_{\Lambda_{1.4}}} \:,
\eeq
being $n$ the number of data pairs, $\bar{L}$ and $\bar{\Lambda}_{1.4}$ the mean values of $L$ and $\Lambda_{1.4}$; and $\sigma_L$ over the data set and $\sigma_{\Lambda_{1.4}}$ their standard deviations. We get a value $r=0.817$, which indicates a weak correlation. 
We note that several EoS lie outside the orange observational band. In particular, we notice that all DDM EoS (blue diamonds), except TW99, are not compatible with the data, as well as all the NLWM EoS (red squares). On the other hand, most of the  Skyrme interactions lie within the orange band, with a few cases incompatible with observations because the predicted $L$ values lie outside the experimental range, and some other are marginally compatible. 
As far as microscopic calculations are concerned, they are all in agreement with GW observations, except the DBHF EoS. 

In the right panel of Fig.~\ref{f:fig3} we report the tidal 
deformability as a function of the radius for a 1.4 solar mass NS, $R_{1.4}$, for the same set of EoS. The observational constraints on $\la_{1.4}$ and $R_{1.4}$ from GW170817 \cite{mergerl} are shown by the orange box. Contrary to the weak $\la_{1.4}-L$ correlation found, we observe a strong quadratic correlation between $\la_{1.4}$ and $R_{1.4}$ the correlation factor being in this case $r=0.986$. This strong correlation was already noticed in Ref.\ \cite{tsang19} using a different set of EoS based again on Skyrme and relativistic mean field models. The behaviour of the microscopic and phenomenological EoS look very similar.


{\section{The neutron skin thickness}
\label{sec:6}}

As stated in the previous Section, correlations between astrophysical observations
and microscopic constraints from nuclear measurements, could help to better understand the properties of nuclear matter. For this purpose,
the limits derived for the tidal deformability in GW170817 could be very valuable and exploited for studying the neutron skin thickness, defined as the difference between the neutron ($R_n$) and proton ($R_p$ ) root-mean-square radii: $\delta R = \sqrt{\langle r_n^2 \rangle} - \sqrt{\langle r_p^2 \rangle}$. It has been shown that this is  
strongly correlated to both $L$ and to the radius of low- mass NS, since the size of a NS and the neutron skin thickness originate both from the pressure of neutron-rich matter, and are sensitive to the same EoS.
As shown by Brown and Typel \cite{skin1,skin2}, and confirmed later by other authors
\cite{stei05,cen09,horo01,skin3,fur02}, the neutron skin thickness calculated in mean field
models with either non-relativistic or relativistic effective interactions, is very sensitive to the
density dependence of the nuclear symmetry energy, and, in particular, to the slope parameter $L$ at normal nuclear
saturation density. Using the Brueckner approach and the several Skyrme forces and relativistic models considered here, the authors of Ref.\ \cite{2009Isaac} made an
estimation of the neutron skin thickness of $^{208}$Pb and $^{132}$Sn, adopting 
the suggestion of Steiner {\it et al.} in Ref.\ \cite{stei05}, where $\delta R$ 
is calculated to lowest order in the diffuseness corrections as $\delta R \sim \sqrt{\frac{3}{5}}t$, being $t$ the thickness of semi-infinite asymmetric
nuclear matter
\begin{equation}
t=\frac{\delta_c}{\rho_0(\delta_c)(1-\delta_c^2)}\frac{E_s}{4\pi r_0^2}
\frac{\int_0^{\rho_0(\delta_c)}\rho^{1/2}[S_0/E_{sym}(\rho)-1][E_{SNM}(\rho)-E_0]^{-1/2} d\rho}
{\int_0^{\rho_0(\delta_c)}\rho^{1/2}[E_{SNM}(\rho)-E_0]^{1/2}d\rho} \ .
\label{eq:t}
\end{equation}

\begin{figure}[!htbp]
\centerline{\includegraphics[scale=0.7]{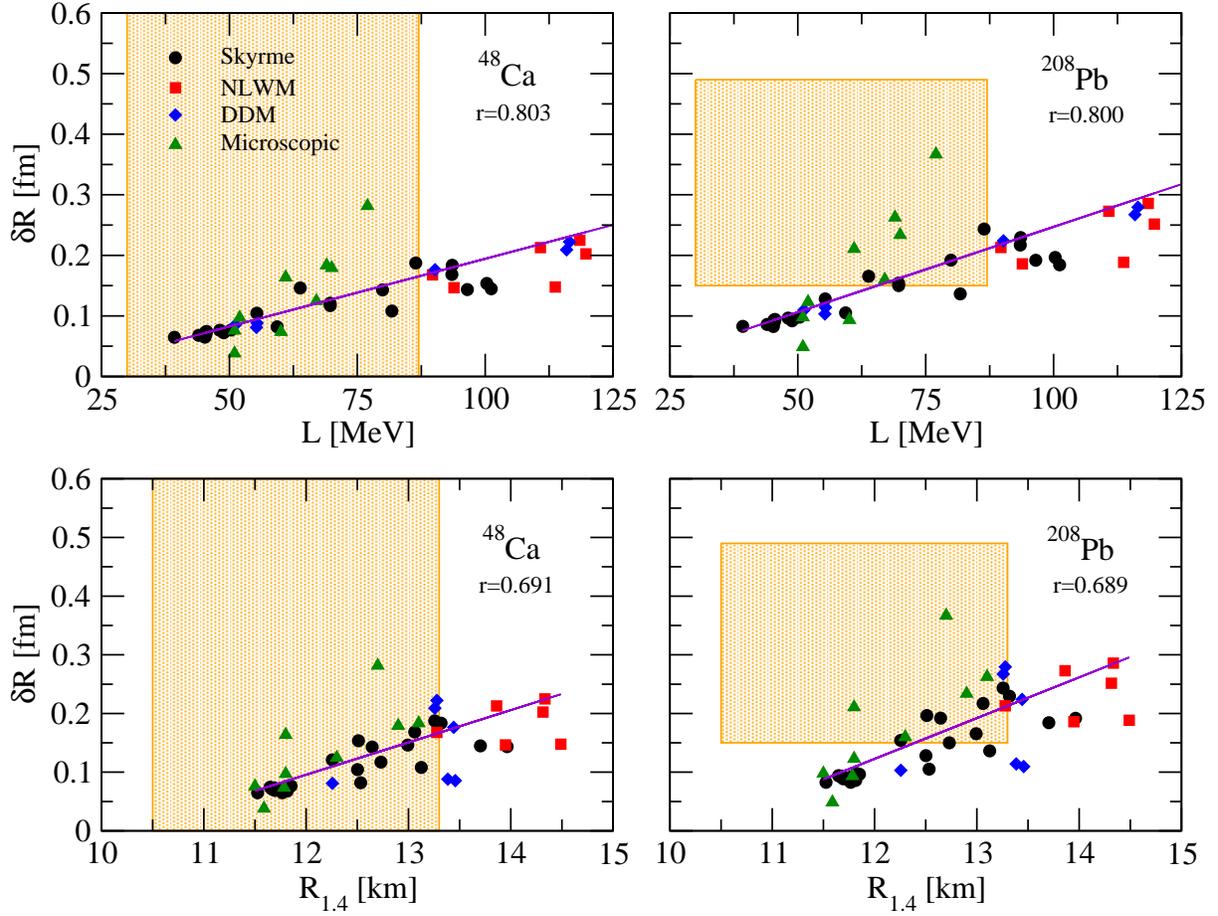}}
\vspace{4mm}
\caption{The neutron skin thickness is displayed vs. the symmetry energy derivative at saturation density L for the different EoS displayed in Table~\ref{t:sat}.
In the left (right) panel calculations are shown for $\rm^{48}Ca$ ($\rm^{208}Pb$). The band on the left panel shows the experimental constraint on $L$, whereas the box on the right one shows in addition the constraint from the PREX experiment \cite{abra12}.  
The violet line indicates a linear fit of the EoS data, Eq.~(\ref{eq:t}). The values of the corresponding correlation factors are also given.
}
\label{f:fig4}
\end{figure}

In the above expression $E_s$ is the surface energy taken from the semi-empirical mass formula equal to $17.23$ MeV,
$r_0$ is obtained from the normalization condition $(4\pi r_0^3/3)(0.16)=1$, and $\delta_c$ is the isospin asymmetry in the center of the nucleus taken as $\delta_c=\delta/2$ according 
to Thomas-Fermi calculations. In this paper, we use the same prescription for the calculation of the neutron skin thickness of $^{208}$Pb and $^{48}$Ca, and we show the results 
in Fig.~\ref{f:fig4}. The orange bands represent the predicted data for $^{48}$Ca (left panels) for which the Calcium Radius Experiment (CREX) has not been run yet \cite{crex}, whereas in the right panels experimental data obtained in
the Lead Radius Experiment (PREX) \cite{abra12} for $^{208}$Pb, $\delta R = 0.33^{+0.16}_{-0.18} \, \rm fm$, are plotted. In the upper panels, results are shown for the neutron skin thickness as a function of the derivative of the symmetry energy $L$. We notice that all the theoretical predictions from phenomenological models and some of the microscopic ones show some correlation between the neutron skin thickness and $L$, as indicated by the linear fits (violet curve) and by the value of the correlation coefficient, $r=0.803$ for $^{48}$Ca ($r=0.800$ for $^{208}$Pb ). Almost all the microscopic EoS turn out to be compatible with the PREX experimental data, whereas some phenomenological models, {\it e.g.} those of the NLWM class, give predictions out of the experimental range.  The linear increase of $\delta$R with L is not surprising since the thickness of the neutron skin in heavy nuclei is determined by the pressure difference between neutrons and protons, which is proportional to the parameter $L$, that is, $P(\rho_0,\delta)\approx L \rho_0 \delta^2/3$. On the other hand, in the lower panels, the neutron skin thickness is displayed as a function of $R_{1.4}$, and in both cases the correlation is very scarce, $r=0.691$ for $^{48}$Ca ($r=0.689$ for $^{208}$Pb ), with a few Skyrme, DDM and all the NLWM EoS incompatible with the observational data. The experimental data from PREX \cite{abra12}, and the upcoming campaigns: PREX-II at Jefferson Lab and the Mainz Radius Experiment (MREX) \cite{mrex} at the future Mainz Energy-Recovering Superconductor Accelerator, can put further strong constraints on the nuclear matter properties, thus selecting the most compatible EoS.

\section{Conclusions}
\label{sec:7}

In this work we have analyzed the existence of possible correlations between NS observables and properties of atomic nuclei. In particular, we have examined correlations of the tidal deformability $\Lambda_{1.4}$ of a 1.4$M_\odot$ NS and the neutron skin thickness $\delta R$ of $^{48}$Ca and $^{208}$Pb with the stellar radius $R_{1.4}$ and the symmetry energy derivative $L$. To such end we have used a large set of different models for the nuclear equation of state, that include microscopic calculations based on the Brueckner--Hartree--Fock and Dirac--Brueckner--Hartree--Fock theories, the variational method and Quantum Monte Carlo techniques, and several phenomenological Skyrme and relativistic mean field models. We have found a strong quadratic correlation between $\Lambda_{1.4}$ and $R_{1.4}$ in agreement with the results of the recent work by Tsang {\it et al.} \cite{tsang19}. On the contrary, we have observed a weaker linear correlation between $\Lambda_{1.4}$ and $L$. Our results have confirmed the existence of a quite linear correlation between the neutron skin thickness of $^{48}$Ca and $^{208}$Pb with $L$, already pointed out by several authors using nonrelativistic and relativistic phenomenological models. A much weaker correlation has been found between $\delta R$ and $R_{1.4}$. The existence of these correlations, predicted by models based on approaches of different nature, suggest that their origin goes beyond the mean field character of the models employed.

To select the most compatible EoS among the ones predicted by the different models considered in this work, we have employed the experimental constraints on $L$ and $\delta R$ together with the observational ones on the mass, radius and tidal deformability imposed by the mass measurement of the millisecond pulsars PSR J1614-2230 \cite{demorest2010} and PSR J0348+0432 \cite{heavy2}, the GW170817 NS merger event \cite{merger,mergerl,mergerx} and the data of the NICER mission \cite{nicer3}. Our results have shown that only five microscopic models (BOB, V18, N93, UIX and DBHF) and four Skyrme forces (SGI, SkMP, SkO and SkO') are simultaneoulsy compatible with the present constraints on $L$ ($30\,\text{MeV} < L < 87\,\text{MeV}$) and the PREX experimental data on the $^{208}$Pb neutron skin thickness.
All the NLWM and DDM models and the majority of the Skyrme forces are excluded by these two experimental constraints. We have also found that almost all the models considered are compatible with the largest masses observed up to now,
$\mmax>2.14^{+0.10}_{-0.09}$ \cite{cromartie} for the object PSR J0740+6620, and PSR J0348+0432 \cite{heavy2},
$M_G=2.01 \pm 0.04 \,M_{\odot}$, and with the upper limit of the maximum mass of about 2.2--2.3$\,\ms$ \cite{shiba17,marga17,rezz18,shiba19} deduced from the analysis of the GW170817 event. Finally, we have seen that
the estimation of the mass ($1.34^{+0.15}_{-0.16}M_\odot$) and equatorial radius ($12.71^{+1.14}_{-1.19}$ km) of the millisecond pulsar PSR J0030+0451 inferred from the Bayesian analysis of the data collected by the NICER mission \cite{nicer3} excludes most of the NLWM EoS considered in this work.

The major experimental, observational and theoretical advances on understanding the nuclear EoS done in the last decades has lead to constrain rather well its isoscalar part. Nevertheless, the isovector part of the nuclear EoS is less well constraint due mainly to our still limited knowledge of the nuclear force and, in particular, of its in-medium modifications and its spin and isospin dependence. Future laboratory experiments being planned in existing or next-generation radioactive ion beam facilities together with further NS observations, particularly a precise simultaneous measurement of the mass and radius of a single object, are fundamental to provide more stringent constraints on the nuclear EoS, and are very much awaited.

\vspace{6pt} 

\authorcontributions{``The authors have read and agreed to the published version of the manuscript.''}

\funding{``This research received no external funding.''}

\acknowledgments{``This work has been supported by the COST Action CA16214 ``PHAROS: The multimessenger physics and astrophysics of neutron stars.''}

\conflictsofinterest{``The authors declare no conflict of interest.''} 

\reftitle{References}


\newcommand{\apjl}{Astrophys. J. Lett.\ }
\newcommand{\apj}{Astrophys. J.\ }
\newcommand{\physrep}{Phys. Rep.\ }
\newcommand{\mnras}{Mon. Not. R. Astron. Soc.\ }
\newcommand{\aap}{Astron. Astrophys.\ }
\newcommand{\prc}{Phys. Rev. C\ }
\newcommand{\prd}{Phys. Rev. D\ }
\newcommand{\prl}{Phys. Rev. Lett.\ }
\newcommand{\nphysa}{Nucl. Phys. A\ }
\newcommand{\plb}{Phys. Lett. B\ }
\newcommand{\epja}{EPJA\ }

\externalbibliography{yes}

\bibliography{corEoS}


\end{document}